\documentclass[showpacs,prd,showkeys,aps]{revtex4}
\usepackage{amsmath}
\usepackage{amssymb}
\newcommand{\beq}{\begin{equation}}
\newcommand{\beqn}{\begin{eqnarray}}
\newcommand{\eeq}{\end{equation}}
\newcommand{\eeqn}{\end{eqnarray}}

\usepackage{braket}
\usepackage{eurosym}
\usepackage{amsfonts}
\usepackage{graphicx}
\usepackage{calrsfs}
\usepackage[usenames,dvipsnames,svgnames,table]{xcolor}

\begin{document}

\title{The Effect of GUP to Massive Vector and Scalar Particles Tunneling From a Warped DGP Gravity Black Hole}

\author{A. \"{O}vg\"{u}n}
\email{ali.ovgun@emu.edu.tr}

\affiliation{Instituto de F\'{\i}sica, Pontificia Universidad Cat\'olica de
Valpara\'{\i}so, Casilla 4950, Valpara\'{\i}so, Chile}
\affiliation{Physics Department, Eastern Mediterranean University, Famagusta,
Northern Cyprus, Turkey}

\author{Kimet Jusufi}
\email{kimet.jusufi@unite.edu.mk}
\affiliation{Physics Department, State University of Tetovo, Ilinden Street nn, 1200, Tetovo,
Macedonia}
\affiliation{Institute of Physics, Faculty of Natural Sciences and Mathematics, Ss. Cyril and Methodius University, Arhimedova 3, 1000 Skopje, Macedonia}

\date{\today }

\begin{abstract}
This paper discusses the effects of the mass and the angular momentum of massive vector and scalar particles on the Hawking temperature manifested under the effects of the generalized uncertainty principle (GUP). In particular, we calculate the Hawking temperature of a black hole in a warped DGP gravity model in the framework of the quantum tunneling method. We use the modified Proca and Klein-Gordon equations previously determined from the GUP Lagrangian in the spacetime background of a warped Dvali-Gabadadze-Porrati (DGP) metric, with the help of Hamilton-Jacobi (HJ) and semiclassical (WKB) approximation methods. We find that as a special case of a warped DGP black hole solution, the Hawking temperature of a Schwarzschild-de Sitter (SdS) black hole can be determined. Furthermore, the Hawking temperature is influenced by the mass and the angular momentum of vector and scalar particles and depends on which of those types of particles is being emitted by the black hole. We conclude that the nonthermal nature of the Hawking spectrum leads to Planck-scale nonthermal correlations, shedding light on the information paradox in black hole evaporation.


\end{abstract}

\maketitle

\section{Introduction}

\label{Introduction} 

Recently, quantum gravity effects have been widely studied in the context of high- and low-energy scales \cite{dgp1,dgp2,dgp3,dgp4,dgp5,dgp6,dgp7,dgp8,dgp9,dgp10,dgp11,dgp12}. One interesting scenario concerns the study of the possible effects of the generalized uncertainty principle (GUP) on Hawking radiation (HR); the effects seem to have interesting implications for the final stage of the black hole evaporation process. In the standard narrative, due to HR, a black hole is thought to lose mass and, eventually, after a very long period of time, to evaporate completely \cite{hawking1,hawking2,hawking3}. However, this scenario may not correctly reflect the final stage of a black hole, since the effects of the GUP modify the Hawking temperature; hence, the black hole may not evaporate completely and may leave remnants \cite{brito1,brito2,abdel1,abdel2,nasser1,nasser2,khalil1,khalil2,khalil3,khalil4,khalil5}.

The amount of HR can be measured by various methods for different types of particles from higher- and lower-dimensional black holes and wormholes \cite{kraus1,kraus2,perkih1,perkih2,perkih3,ang,sri1,sri2,vanzo,mann0,mann1,mann2,mann3,xiang,kruglov1,sakalli1,sakalli2,sakalli3,kimet,J1,S1,O1,S2,S3,O2,O3,S4,Kepl}. In addition to the mass, charge, and angular momentum of a black hole, the Hawking temperature is shown to depend on the nature of the particles emitted by the black hole, when the effects of the GUP are taken into account. In \cite{qgv1} and \cite{qgv2}, for the cases of scalar and Dirac particles, the corrected Hawking temperature was shown to depend on the particles’ mass and angular momentum. Later, in Ref. \cite{xiang2}, a similar result was found for the corrected Hawking temperature associated with the tunneling of massive vector particles and related to the mass and angular momentum of particles emitted by the black hole. This is quite an interesting result, since it could be related to the information loss paradox. 

On the other hand, it is suggested that living on a brane embedded in extradimensional space causes the weakness of gravity \cite{dvali1,dvali2}. By compactifying extradimensional space, true four-dimensional (4D) gravity may be produced at large distances because of the finite volume of the extradimensional space. Moreover, one can produce the same effect by forcing extradimensional space to warp \cite{dvali4}, if the extra space has a finite size. Then, one modifies 4D gravity with flat, infinitely sized extra dimensions to illuminate the problems of supersymmetry breakage and the cosmological constant\cite{dvali3}. In Dvali-Gabadadze-Porrati (DGP) gravity, a mechanism is proposed by which 4D Newtonian gravity is produced on a 3-brane embedded in 5D Minkowski space \cite{dvali3}. It is noted that in the DGP model, the potential becomes 4D at short distances but behaves like 5D at large distances \cite{dvali3}.

Theories of gravity in more than four dimensions have recently attracted much interest — for example, the black hole solution recently found in a warped DGP brane model \cite{Bh}. Inspired by this work, we aim to extend the quantum tunneling method to massive vector and scalar particles under the influence of the GUP effects from a warped DGP brane-model black hole (WDGPBBH). We shall consider the GUP-modified Proca and Klein-Gordon equations (published recently in the literature) in the spacetime of a WDGPBBH. 

This paper is organized as follows. In Section 2, we review the warped DGP brane-model solution published recently \cite{Bh}. In Section 3, we review the GUP-corrected Lagrangian of massive vector fields and solve the GUP-modified Proca equation \cite{xiang2} to find the corrected Hawking temperature of the warped DGP black hole. In Section 4, we solve the GUP-modified Klein-Gordon equation and determine the corrected Hawking temperature. In Section 5, we discuss our results.

\section{Black Holes in DGP Gravity}

We consider a 5-D bulk spacetime with a single 4-D brane, on which
gravity is confined, and derive the effective 4-D gravitational equations. First, one locates the 4-D brane $(M,g_{\mu\nu})$ at a hypersurface
(${ B}(X^{A})=0$) in the 5-D bulk spacetime $({ M},{}^{(5)}g_{AB})$. It is noted that the coordinates are  $X^{A}~(A=0,1,2,3,5)$. The action of the brane world is given as follows:
\begin{eqnarray}
S=S_{{\rm bulk}}+S_{{\rm brane}},\label{action}
\end{eqnarray}
where 
\begin{eqnarray}
S_{{\rm bulk}}=\int_{{\cal M}}d^{5}X\sqrt{-^{(5)}g}\left[\frac{1}{2\kappa_{5}^{2}}{}^{(5)}R+{}^{(5)}L_{{\rm m}}\right],\label{bulk_action}
\end{eqnarray}
and 
\begin{eqnarray}
S_{{\rm brane}}=\int_{M}d^{4}x\sqrt{-g}\left[\frac{1}{\kappa_{5}^{2}}K^{\pm}+L_{{\rm brane}}(g_{\alpha\beta},\psi)\right].\label{brane_action}
\end{eqnarray}
Note that $\kappa_{5}^{2}$ is the 5-D gravitational constant. Moreover $^{(5)}R$ and
$^{(5)}L_{{\rm m}}$ stand for the 5-D scalar curvature and the matter in the bulk, respectively. It is defined the induced 4-D coordinates on the brane as $x^{\mu}~(\mu=0,1,2,3)$ where the trace of the trace of extrinsic
curvature is $K^{\pm}$. On the other hand, the effective 4-D Lagrangian is  $L_{{\rm brane}}(g_{\alpha\beta},\psi)$
with a generic functional
of the brane metric $g_{\alpha\beta}$ and matter fields $\psi$. 

The 5-D Einstein equations in the bulk are derived from the action as follows: 
\begin{eqnarray}
^{(5)}G_{AB}=\kappa_{5}^{2}\,\,\left[\,{}^{(5)}T_{AB}+\tau_{AB}\,\delta({ B})\,\right]\,,\label{5dEinstein}
\end{eqnarray}
where $\tau_{\mu\nu}$
is the ``effective\char`\"{} energy-momentum tensor localized on
the brane, $^{(5)}T_{AB}$ is the energy-momentum tensor of bulk matter fields and  $\delta({ B})$ is the localization of brane contributions:
\begin{eqnarray}
^{(5)}T_{AB} & \equiv & -2\frac{\delta{}^{(5)}\!L_{{\rm m}}}{\delta{}^{(5)}g^{AB}}+{}^{(5)}g_{AB}{}^{(5)}\!L_{{\rm m}}\, ,\label{em_tensor_of_bulk}
\end{eqnarray}

\begin{eqnarray}
\tau_{\mu\nu}\equiv-2\frac{\delta L_{{\rm brane}}}{\delta g^{\mu\nu}}+g_{\mu\nu}L_{{\rm brane}}\,.\label{em_tensor_of_brane}
\end{eqnarray}

As it is solved the field equations and  derived the black hole solution in 4-D brane world \cite{Bh}, the black hole solution on the brane is found by:

\begin{equation}
ds^{2}=-f(r)dt^{2}+f(r)^{-1}dr^{2}+r^{2}\left(d\theta^{2}+\text{sin}^{2}\theta d\phi^{2}\right),
\end{equation}
with $f(r)=1-\frac{r_{M}}{r}-\frac{r^{2}}{r_{c}^{2}}.$ It is noted
that it reduces to the asymptotically de Sitter space for $r_{c}=\sqrt{\frac{3}{\Lambda}}$
, $r=\frac{r_{c}}{2}$ and $r_{M}=0$. Note that also it reduces to
Schwarzchild black hole for $r_{c}\rightarrow\infty$ and $r=r_{M}=2M$.

\section{Massive vector particles tunneling from WDGPBBH by the GUP}

\label{Section2} In this section, firstly, we study the massive vector particles
tunneling from WDGPBBH by the GUP using the method in Ref. \cite{xiang2}. Starting from the GUP-corrected Lagrangian of massive vector field $\mathfrak{\varPsi}_{\mu}$ given by \cite{xiang2}
\begin{equation}
\mathcal{L}^{GUP}=-\frac{1}{2}\left(\mathcal{D}_{\mu}\mathfrak{\varPsi}_{\nu}-\mathcal{D}_{\nu}\mathfrak{\varPsi}_{\mu}\right)\left(\mathcal{D}^{\mu}\mathfrak{\varPsi}^{\nu}-\mathcal{D}^{\nu}\mathfrak{\varPsi}^{\mu}\right)-\frac{m_W^2}{\hbar^2}\mathfrak{\varPsi}_{\mu}\mathfrak{\varPsi}^{\mu}.
\end{equation}

The modified field equation for massive
bosons in the case of uncharged bosons $W$, is given as follows \cite{xiang2}:
\begin{equation}
\partial_{\mu}\left(\sqrt{-g}\mathfrak{\varPsi}^{\mu\nu}\right)-\sqrt{-g}\frac{m_W^{2}}{\hbar^{2}}\mathfrak{\varPsi}^{\nu}+\beta\hbar^{2}\partial_{0}\partial_{0}\partial_{0}\left(\sqrt{-g}g^{00}\mathfrak{\varPsi}^{0\nu}\right)-\beta\hbar^{2}\partial_{i}\partial_{i}\partial_{i}\left(\sqrt{-g}g^{ii}\mathfrak{\varPsi}^{i\nu}\right)=0,\label{Fe2}
\end{equation}
where the GUP modified antisymmetric tensor $\mathfrak{\varPsi}_{\mu\nu}$ is given as $\mathfrak{\varPsi}_{\mu\nu}=(1-\beta \hbar^2 \partial_{\mu}^2)\,\partial_{\mu}\mathfrak{\varPsi}_{\nu}-(1-\beta \hbar^2 \partial_{\nu}^2)\,\partial_{\nu}\mathfrak{\varPsi}_{\mu} $. We note that in the case of the modified tensor $\mathfrak{\varPsi}_{i\mu}$, the Latin indices refer to spatial components i.e., $i=1,2,3$, while in the case of $\mathfrak{\varPsi}_{0\mu}$, the $0$ denotes the time coordinate. Furthermore $\beta$ can be written in terms of the minimal length $(M_{f})$ as, $\beta = 1/(3M_{f}^2 )$. On the other hand $m$ is the mass of the vector particle.
The metric of the spherically symmetric static spacetime WDGPBBH is
given by 

\begin{equation}
ds^{2}=-f(r)dt^{2}+f(r)^{-1}dr^{2}+r^{2}\left(d\theta^{2}+\text{sin}^{2}\theta d\phi^{2}\right),\label{RNmetric}
\end{equation}
where 
\begin{equation}
f(r)=1-\frac{r_{M}}{r}-\frac{r^{2}}{r_{c}^{2}}.\label{RNmetricfu}
\end{equation}

According to the WKB approximation, the vector field $\mathfrak{\varPsi}_{\mu}$, has the form of
\begin{equation}
\mathfrak{\varPsi}_{\mu}=c_{\mu}(t,r,\theta,\phi){\rm exp}\left[\frac{i}{\hbar}I(t,r,\theta,\phi)\right],\label{WKBW}
\end{equation}
where $I$ is defined as 
\begin{equation}
I(t,r,\theta,\phi)=I_{0}(t,r,\theta,\phi)+\hbar I_{1}(t,r,\theta,\phi)+\hbar^{2}I_{2}(t,r,\theta,\phi)+\cdots.\label{S0123}
\end{equation}
By substituting Eqs~\eqref{WKBW},~\eqref{S0123}, and the WDGPBBH
metric~\eqref{RNmetric} into Eq.~\eqref{Fe2}, and keeping
only the lowest order in $\hbar$, we obtain the equations for the
coefficients $c_{\mu}$:
\begin{eqnarray}
f(r)\left[c_{0}(\partial_{r}I_{0})^{2}\mathcal{A}_{1}^{2}-c_{1}(\partial_{r}I_{0})(\partial_{t}I_{0})\mathcal{A}_{1}\mathcal{A}_{0}\right]\nonumber \\
+\frac{1}{r^{2}}\left[c_{0}(\partial_{\theta}I_{0})^{2}\mathcal{A}_{2}^{2}-c_{2}(\partial_{\theta}I_{0})(\partial_{t}I_{0})\mathcal{A}_{2}\mathcal{A}_{0}\right]\nonumber \\
+\frac{1}{r^{2}{\rm sin}^{2}\theta}\left[c_{0}(\partial_{\phi}I_{0})^{2}\mathcal{A}_{3}^{2}-c_{3}(\partial_{\phi}I_{0})(\partial_{t}I_{0})\mathcal{A}_{3}\mathcal{A}_{0}\right]+c_{0}m_{W}^{2}=0,
\end{eqnarray}
\begin{eqnarray}
-\frac{1}{f(r)}\left[c_{1}(\partial_{t}I_{0})^{2}\mathcal{A}_{0}^{2}-c_{0}(\partial_{t}I_{0})(\partial_{r}I_{0})\mathcal{A}_{0}\mathcal{A}_{1}\right]\nonumber \\
+\frac{1}{r^{2}}\left[c_{1}(\partial_{\theta}I_{0})^{2}\mathcal{A}_{2}^{2}-c_{2}(\partial_{\theta}I_{0})(\partial_{r}I_{0})\mathcal{A}_{2}\mathcal{A}_{1}\right]\nonumber \\
+\frac{1}{r^{2}{\rm sin}^{2}\theta}\left[c_{1}(\partial_{\phi}I_{0})^{2}\mathcal{A}_{3}^{2}-c_{3}(\partial_{\phi}I_{0})(\partial_{r}I_{0})A_{3}\mathcal{A}_{1}\right]+c_{1}m_{W}^{2}=0,
\end{eqnarray}
\begin{eqnarray}
-\frac{1}{f(r)}\left[c_{2}(\partial_{t}I_{0})^{2}\mathcal{A}_{0}^{2}-c_{0}(\partial_{t}I_{0})(\partial_{\theta}I_{0})\mathcal{A}_{0}\mathcal{A}_{2}\right]\nonumber \\
+f(r)\left[c_{2}(\partial_{r}I_{0})^{2}\mathcal{A}_{1}^{2}-c_{1}(\partial_{r}I_{0})(\partial_{\theta}I_{0})\mathcal{A}_{1}\mathcal{A}_{2}\right]\nonumber \\
+\frac{1}{r^{2}{\rm sin}^{2}\theta}\left[c_{2}(\partial_{\phi}I_{0})^{2}\mathcal{A}_{3}^{2}-c_{3}(\partial_{\phi}I_{0})(\partial_{\theta}I_{0})\mathcal{A}_{3}\mathcal{A}_{2}\right]+c_{2}m_{W}^{2}=0,
\end{eqnarray}
\begin{eqnarray}
-\frac{1}{f(r)}\left[c_{3}(\partial_{t}I_{0})^{2}\mathcal{A}_{0}^{2}-c_{0}(\partial_{t}I_{0})(\partial_{\phi}I_{0})\mathcal{A}_{0}\mathcal{A}_{3}\right]\nonumber \\
+f(r)\left[c_{3}(\partial_{r}I_{0})^{2}\mathcal{A}_{1}^{2}-c_{1}(\partial_{r}I_{0})(\partial_{\phi}I_{0})\mathcal{A}_{1}\mathcal{A}_{3}\right]\nonumber \\
+\frac{1}{r^{2}}\left[c_{3}(\partial_{\theta}I_{0})^{2}\mathcal{A}_{2}^{2}-c_{2}(\partial_{\theta}I_{0})(\partial_{\phi}I_{0})\mathcal{A}_{2}\mathcal{A}_{3}\right]+c_{3}m_{W}^{2}=0,
\end{eqnarray}
where the $\mathcal{A}_{\mu}$s are defined as 
\begin{eqnarray}
 &  & \mathcal{A}_{0}=1+\beta\frac{1}{f(r)}(\partial_{t}S_{0})^{2},\ \mathcal{A}_{1}=1+\beta f(r)(\partial_{r}S_{0})^{2},\nonumber \\
 &  & \mathcal{A}_{2}=1+\beta\frac{1}{r^{2}}(\partial_{\theta}S_{0})^{2},\ \mathcal{A}_{3}=1+\beta\frac{1}{r^{2}{\rm sin}^{2}\theta}(\partial_{\phi}S_{0})^{2}.
\end{eqnarray}
Considering the property of WDGPBBH spacetime and the question that
we aim to address, then following the standard process, we separate
the variables 
\begin{equation}
I_{0}=-Et+R(r)+\Theta(\theta,\phi),\label{RNs0fj}
\end{equation}
where $E$ is the energy of the emitted vector particles. Then  a matrix equation is obtained as
\begin{equation}
\Sigma(c_{0},c_{1},c_{2},c_{3})^{T}=0,\label{matrixeq}
\end{equation}
where $\Sigma$ is a 4$\times$4 matrix, the elements of which are
\begin{eqnarray}
 &  & \Sigma_{11}=f(r)R'^{2}\mathcal{A}_{1}^{2}+\frac{{J_{\theta}}^{2}}{r^{2}}\mathcal{A}_{2}^{2}+\frac{{J_{\phi}}^{2}}{r^{2}{\rm sin}^{2}\theta}\mathcal{A}_{3}^{2}+m_{W}^{2},\ \Sigma_{12}=-f(r)R'(-E)\mathcal{A}_{1}\mathcal{A}_{0},\nonumber \\
 &  & \Sigma_{13}=-\frac{J_{\theta}(-E)}{r^{2}}\mathcal{A}_{2}\mathcal{A}_{0},\ \Sigma_{14}=-\frac{J_{\phi}(-E)}{r^{2}{\rm sin}^{2}\theta}\mathcal{A}_{3}\mathcal{A}_{0},\nonumber \\
 &  & \Sigma_{21}=\frac{(-E)R'}{f(r)}\mathcal{A}_{0}\mathcal{A}_{1},\ \Sigma_{22}=-\frac{(-E)^{2}}{f(r)}\mathcal{A}_{0}^{2}+\frac{{J_{\theta}}^{2}}{r^{2}}\mathcal{A}_{2}^{2}+\frac{J_{\phi}^{2}}{r^{2}{\rm sin}^{2}\theta}\mathcal{A}_{3}^{2}+m_{W}^{2},\nonumber \\
 &  & \Sigma_{23}=-\frac{J_{\theta}R'}{r^{2}}\mathcal{A}_{2}\mathcal{A}_{1},\ \Sigma_{24}=-\frac{J_{\phi}R'}{r^{2}{\rm sin}^{2}\theta}\mathcal{A}_{3}A_{1},\nonumber \\
 &  & \Sigma_{31}=\frac{-EJ_{\theta}}{f(r)}\mathcal{A}_{0}\mathcal{A}_{2},\ \Sigma_{32}=-f(r)R'J_{\theta}\mathcal{A}_{1}\mathcal{A}_{2},\\
 &  & \Sigma_{33}=-\frac{(-E)^{2}}{f(r)}\mathcal{A}_{0}^{2}+f(r)R'^{2}\mathcal{A}_{1}^{2}+\frac{J_{\phi}^{2}}{r^{2}{\rm sin}^{2}\theta}\mathcal{A}_{3}^{2}+m_{W}^{2},\ \Sigma_{34}=-\frac{J_{\theta}J_{\phi}}{r^{2}{\rm sin}^{2}\theta}\mathcal{A}_{3}\mathcal{A}_{2},\nonumber \\
 &  & \Sigma_{41}=\frac{(-E)J_{\phi}}{f(r)}\mathcal{A}_{0}\mathcal{A}_{3},\ \Sigma_{42}=-f(r)R'J_{\phi}\mathcal{A}_{1}\mathcal{A}_{3},\nonumber \\
 &  & \Sigma_{43}=-\frac{J_{\theta}J_{\phi}}{r^{2}}\mathcal{A}_{2}\mathcal{A}_{3},\ \Sigma_{44}=-\frac{(-E)^{2}}{f(r)}\mathcal{A}_{0}^{2}+f(r)R'^{2}A_{1}^{2}+\frac{J_{\theta}^{2}}{r^{2}}\mathcal{A}_{2}^{2}+m_{W}^{2},\nonumber 
\end{eqnarray}
where $R'=\partial_{r}R$, $J_{\theta}=\partial_{\theta}\Theta$ and
$J_{\phi}=\partial_{\phi}\Theta$.

Eq.~\eqref{matrixeq} has a nontrivial solution if the determinant
of the matrix $\Sigma$ equals zero. By neglecting the higher order terms of $\beta$ and solving ${\rm det}\Sigma=0$,
we obtain the solution to the derivative of the radial action 
\begin{equation}
\partial_{r}R_{\pm}=\pm\sqrt{-\frac{m^{2}}{f(r)}+\frac{(E)^{2}}{f(r)^{2})}-\frac{J_{\theta}^{2}+J_{\phi}^{2}{\rm csc}^{2}\theta}{f(r)r^{2}}}\left(1+\frac{\mathcal{T}_{1}}{\mathcal{T}_{2}}\beta\right),\label{wpr}
\end{equation}
where 
\begin{eqnarray}
\mathcal{T}_{1} & = & -3f(r)m^{4}r^{2}+6m^{2}r^{2}(E)^{2}-6f(r)m^{2}(J_{\theta}^{2}+J_{\phi}^{2}{\rm csc}^{2}\theta)-\frac{6f(r)J_{\theta}^{4}}{r^{2}}\nonumber \\
 & + & 6(E)^{2}(J_{\theta}^{2}+J_{\phi}^{2}{\rm csc}^{2}\theta)-\frac{7f(r)J_{\theta}^{2}J_{\phi}^{2}{\rm csc}^{2}\theta}{r^{2}}-\frac{3f(r)J_{\theta}^{4}J_{\phi}^{2}{\rm csc}^{2}\theta}{2m^{2}r^{4}}\nonumber \\
 & - & \frac{5f(r)J_{\phi}^{4}{\rm csc}^{4}\theta}{r^{2}}+\frac{3f(r)J_{\theta}^{2}J_{\phi}^{4}{\rm csc}^{4}\theta}{2m^{2}r^{4}},\\
\mathcal{T}_{2} & = & -f(r)m^{2}r^{2}+r^{2}(E)^{2}-f(r)(J_{\theta}^{2}+J_{\phi}^{2}{\rm csc}^{2}\theta).
\end{eqnarray}

Integrating Eq.~\eqref{wpr} around the pole at the outer horizon
yields the solution of the radial action. The particle's tunneling
rate is determined by the imaginary part of the action, 
\begin{eqnarray}
\text{Im} R_{\pm}(r) & = & \pm \text{Im} \int dr\sqrt{-\frac{m^{2}}{f(r)}+\frac{(E)^{2}}{f(r)^{2}}-\frac{J_{\theta}^{2}+J_{\phi}^{2}{\rm csc}^{2}\theta}{f(r)r^{2}}}\left(1+\frac{\mathcal{T}_{1}}{\mathcal{T}_{2}}\beta\right).\label{in1}
\end{eqnarray}

In order to solve this integral we must write the metric \eqref{RNmetric} near the event horizon, thus following \cite{rahman} and solving $r^{3}+2Mr_c^{2}-rr_c^{2}=0$, one gets the black hole event horizon $r_{h}$, and the cosmological horizon $\mathfrak{r}_{cos}$, given by
\begin{equation}
r_{h}=\frac{2M}{\sqrt{3\xi}}\cos\frac{\pi+\psi}{3},\label{5}
\end{equation}
\begin{equation}
\mathfrak{r}_{cos}=\frac{2M}{\sqrt{3\xi}}\cos\frac{\pi-\psi}{3},
\end{equation}
where
\begin{equation}
\psi=\cos^{-1}(3\sqrt{3\xi}).
\end{equation}

In which $\xi=M^{2}/r_c^{2}$, and belongs to the interval $0<\xi<1/27$. Moreover by expanding $r_ {h} $ in terms of $M$ in the interval $\xi<1/27$, gives (see, e.g., \cite{rahman})
\begin{equation}
r_{h}=2M\left(1+\frac{4M^{2}}{r_c^{2}}+\cdots\right),\label{rh}
\end{equation}

Let us define $\Delta(r)=r^2-2Mr-r^4/r_c^2$,  so that 
\begin{equation}
\Delta_{,r} \, (r_{h})=\frac{d\Delta }{dr}|_{r_h}=2\left(r_{h}-M-\frac{2 r_h^3}{r_c^2} \right).
\end{equation}

In other words, the metric \eqref{RNmetric} can be written as 
\begin{equation}
ds^2=-\frac{\Delta_{,r} \,(r_{h})}{r_h{^2}}(r-r_h)dt^2+\frac{r_h{^2} }{\Delta_{,r} \, (r_{h}) (r-r_h)}dr^2+r_h^2 d\theta^2+r_h^2 \sin^2 \theta d\phi^2,
\end{equation}
in which $r_{h}$ is given by Eq. \eqref{rh}. From the last metric the following identification for $f(r_h)$ can be written 
\begin{equation}
f(r_h)\approx \frac{\Delta_{,r}  \,(r_{h})}{r_h{^2}}(r-r_h).
\end{equation}

Solving the integral \eqref{in1}, we find the following result for the radial part
\begin{equation}
 \text{Im} R_{\pm}(r) = \pm i \pi\frac{r_{h}^{2}}{\Delta_{,r} \, (r_{h})}(E)\times\left(1+\beta\Xi\right),
\end{equation}
where $\Xi=6m^{2}+\frac{6}{r_{h}^{2}}\left(J_{\theta}^{2}+J_{\phi}^{2}{\rm csc}^{2}\theta\right)$.

It is quite clear that $\Xi>0$. We note that $R_{+}$ represents
the radial function for the outgoing particles and $R_{-}$ is for
the ingoing particles. Thus, the tunneling rate of $W$ bosons
near the event horizon is 
\begin{eqnarray}
\Gamma & = & \frac{P_{outgoing}}{P_{ingoing}}=\frac{{\rm exp}\left[-\frac{2}{\hbar}({\rm Im}R_{+}+{\rm Im}\Theta)\right]}{{\rm exp}\left[-\frac{2}{\hbar}({\rm Im}R_{-}+{\rm Im}\Theta)\right]}={\rm exp}\left[-\frac{4}{\hbar}{\rm Im}R_{+}\right]\nonumber \\
 & = & \exp\left[-\frac{4\pi}{\hbar}\frac{r_{h}^{2}}{\Delta_{,r}\, (r_{h})}(E)\times\left(1+\beta\Xi\right)\right].
\end{eqnarray}
If we set $\hbar=1$, then the effective Hawking temperature is deduced
as 
\begin{eqnarray}
T_{e-H}=\frac{\Delta_{,r} \, (r_{h})}{4\pi r_{h}^{2}\left(1+\beta\Xi\right)}=T_{0}\left(1-\beta\Xi\right),\label{effectiveHT}
\end{eqnarray}
where $T_{0}=\frac{\Delta_{,r} \, (r_{h})}{4\pi r_{h}^{2}}$ is the original
Hawking temperature of a Warped DGP gravity black hole which is similar to the Schwarzschild-de Sitter (SdS) black hole \cite{rahman}. Moreover we recover the Hawking temperature for the Schwarzchild black hole in the limit $r_{c}\rightarrow\infty$ and $r_{h}=2M$. From Eq.~\eqref{effectiveHT}, it can be inferred that the corrected temperature relies on the quantum
numbers (mass and angular momentum) of the emitted vector bosons.
Moreover, the quantum effects explicitly counteract the temperature
increases during evaporation, which will cancels it out at some point.
Naturally, black hole remnants will be left.

\section{Tunneling of massive SCALAR PARTICLES WITH GUP}

We turn our attention now to the case of scalar particles under GUP effects by WDGPBBH black hole. We can easily incorporate GUP effects into the massive Klein-Gordon equation by using
the modified operators of position and momentum. In leading order of $\beta$, this leads to the following equation \cite{qgv1}
\begin{equation}
-(i \hbar )^2 \partial^t \partial_t \Phi =\left[(i \hbar )^2 \partial^i \partial_i+m^2 \right]\left[1-2\beta \left( (i \hbar )^2 \partial^i \partial_i+m^2  \right)  \right] \Phi.
\end{equation}

Furthermore, after we choose the following ansatz for the scalar field $\Phi$:

\begin{equation}
\Phi(t,r,\theta,\phi)=\left[\frac{i}{\hbar}I(t,r,\theta,\phi)\right],
\end{equation}
and considering only the lowest order terms in $\hbar$, the following equation can be found \cite{qgv1}

\begin{eqnarray}
&&\frac{1}{f} (\partial_t I_0)^2 = \left[\left( f(\partial_r I_0)^2+\frac{1}{r^2}(\partial_\theta I_0)^2 +\frac{1}{r^2 \sin^2 \theta } (\partial_\phi I_0)^2 \right) +m^2 \right] \times \\ 
&& \left[1-2 \beta \left(  f(\partial_r I_0)^2+\frac{1}{r^2}(\partial_\theta I_0)^2 +\frac{1}{r^2 \sin^2 \theta } (\partial_\phi I_0)^2  +m^2 \right)   \right].
\end{eqnarray}

Now let us carry out the separation of variables 
\begin{equation}
I_{0}=-Et+R(r,\theta)+j \phi,
\end{equation}
then we can fix the angle $\theta=\theta_0$, which yields to the following equation
\begin{equation}
A \,(\partial_r R)^4+B \,(\partial_r R)^2+C=0,\label{40}
\end{equation}
where 
\begin{eqnarray}
A &=& -2 \beta f^2(r),\\
B &= & f(r) \left( 1-\frac{4 \beta j^2 }{r^2 \sin^2 \theta}-4 \beta m^2 \right),\\
C &= &  m^2 +\frac{j^2}{r^2 \sin^2 \theta}-\frac{2 \beta j^4}{r^4 \sin^4 \theta}-\frac{4\beta m^2 j^2 }{r^2 \sin^2 \theta}-2\beta m^4 -\frac{E^2}{f(r)}.
\end{eqnarray}

After some algebraic manipulations, Eq. \eqref{40} yields the following integral for the radial part for the wave equation
\begin{eqnarray}
R(r)=\pm \int \frac{\sqrt{E^2-f(r)\left[m^2+\frac{j^2 }{r^2 \sin^2 \theta }-2 \beta \left( \frac{j^4}{r^4 \sin^4 \theta }+\frac{2 m^2 j^2}{r^2 \sin^2 \theta}+m^4\right) \right]  }}{f(r)} \left[1+\beta\left(m^2+\frac{E^2}{f(r)}+\frac{ j^2}{r^2 \sin^2 \theta} \right) \right] dr.
\end{eqnarray}

Solving this integral and neglecting higher order terms in $\beta$,  we find the following result near the black hole event horizon
\begin{equation}
\text{Im} \,R(r_{h})=\pm i \pi \frac{r_{h}^2\,E}{\Delta_{,r} \, (r_{h})}\left(1+\beta \zeta \right),
\end{equation}
where
\begin{equation}
\zeta(r_{h})=\frac{m^2}{2}+\frac{j^2 \csc^2 \theta }{2 \,r_{h}^2}.
\end{equation}

In this section we shall recover the Hawking temperature by considering the ambiguity of a factor two associated with the solution known as a factor of two problem (see, for example \cite{Akhmedova0,Akhmedova1,borun,Akhmedova2,Akhmedova3,Akhmedova4,yale,singleton1,singleton2}). We should consider the canonical invariance under canonical transformations given by  $\oint p_{r}\mathrm{d}r=\int p_{r}^{+}\mathrm{d}r-\int p_{r}^{-}\mathrm{d}r$, where $p_{r}^{\pm}=\pm\partial_{r}R$, then one should compute the spatial contribution to the tunneling as well as temporal part contribution.  The spatial contribution reads
\begin{eqnarray}\nonumber
\Gamma_{spatial}&\propto &\exp\left(-\frac{1}{\hbar}\text{Im} \oint p_{r} \mathrm{d}r\right)\\\nonumber
&=&\exp\left[-\frac{1}{\hbar} \text{Im} \left(\int p_{r}^{+}\mathrm{d}r-\int p_{r}^{-}\mathrm{d}r\right) \right]\\
&=& \exp \left[-\frac{2\pi}{\hbar} \frac{r_{h}^2\,E}{\Delta_{,r} \, (r_{h})}\left(1+\beta \zeta \right)\right].
\end{eqnarray}

The temporal part comes due to the connection of the interior region and the exterior region of the black hole. Introducing $t\to t -i\pi/(2\kappa) $ which suggests that  Im($E\Delta t^{out,in})=-E\pi/(2\kappa)$. Thus, the total temporal contribution can be calculated as
\begin{eqnarray}\nonumber
\Gamma_{temp.}&\propto &\exp\left[\frac{1}{\hbar}\left(\text{Im}(E\Delta t^{out})+\text{Im}(E\Delta t^{in}\right)\right]\\
&=&\exp \left[-\frac{2\pi}{\hbar} \frac{r_{h}^2\,E}{\Delta_{,r} \, (r_{h})}\left(1+\beta \zeta \right)\right].
\end{eqnarray}

The total tunneling rate at the horizon reads
\begin{eqnarray}\nonumber
\Gamma &=& \exp \Big[\frac{1}{\hbar}\Big(\text{Im}(E\Delta t^{out})+\text{Im}(E\Delta t^{in})
-\text{Im}\oint p_{r} dr \Big)\Big]\\
&=& \exp \left[-\frac{4\pi}{\hbar} \frac{r_{h}^2\,E}{\Delta_{,r} \, (r_{h})}\left(1+\beta \zeta \right)\right].
\end{eqnarray}

If we set $\hbar=1$, and make use of the Boltzmann equation $\Gamma_B=\exp(-E/T_{H})$, the effective Hawking temperature is calculated as
\begin{eqnarray}
T_{e-H}=\frac{\Delta_{,r}  (r_{h})}{4\pi r_{h}^{2}\left(1+\beta\zeta \right)}=T_{0}\left(1-\beta\zeta  \right).
\end{eqnarray}

Hence, we found that quantum gravity corrected Hawking temperature depends on the particle's mass and angular momentum associated with the scalar particles. Although this expression looks similar to the corrected Hawking temperature for vector particles Eq. \eqref{effectiveHT}, the relation to mass and angular momentum are not completely identical in these results.

\section{Discussion and conclusion}

We have studied the effect of quantum gravity on the massive vector particles tunneling from a WDGPBBH and derived their tunneling rates. We have pointed out that the radiation spectra is not purely thermal and that the GUP parameter $\beta$ affects the tunneling rate. We have shown that the Hawking temperature of a warped DGP-gravity black hole is identical to the SdS black hole temperature. We see that when quantum gravity effects are introduced, the nature of the particles emitted by the black hole plays an important role. In other words, due to quantum gravity effects, the remnants are more likely to form during the evaporation of vector particles than during the evaporation of scalar particles — i.e., $\Xi (r_{h}) >\zeta (r_{h}) $. Finally, the difference in the Hawking temperature due to the type of particles tunneled from the black hole could shed light on the information loss paradox in the near future.

\begin{acknowledgments}
This work was supported by the Chilean FONDECYT Grant No. 3170035 (A\"{O}).
\end{acknowledgments}

\end{document}